# Stability of Tsallis entropy and instabilities of Rényi and normalized Tsallis entropies: A basis for $q$-exponential distributions

Sumiyoshi Abe

*Institute of Physics*, *University of Tsukuba*,

*Ibaraki 305-8571*, *Japan*

The $q$-exponential distributions, which are generalizations of the Zipf-Mandelbrot power-law distribution, are frequently encountered in complex systems at their stationary states. From the viewpoint of the principle of maximum entropy, they can apparently be derived from three different generalized entropies: the Rényi entropy, the Tsallis entropy, and the normalized Tsallis entropy. Accordingly, mere fittings of observed data by the $q$-exponential distributions do not lead to identification of the correct physical entropy. Here, stabilities of these entropies, i.e., their behaviors under arbitrary small deformation of a distribution, are examined. It is shown that, among the three, the Tsallis entropy is stable and can provide an entropic basis for the $q$-exponential distributions, whereas the others are unstable and cannot represent any experimentally observable quantities.

arXiv:cond-mat/0206078 v3  10 Jun 2002

## I. INTRODUCTION

It is known [1-3] that there are a number of complex systems whose statistical properties at the stationary states are well described by the *q*-exponential distributions, which are generalizations of the Zipf-Mandelbrot power-law distribution [4]. The *q*-exponential distributions are anomalous distributions from the viewpoint of conventional statistical mechanics characterized by Boltzmann's exponential factor. Since so frequently observed in nature, it is of importance to develop bases for such distributions. In this context, we wish to mention that quite recently the *q*-exponential factor has been obtained for the logistic map at the edge of chaos by the renormalization group method as well as by the Pesin equality for the generalized Kolmogorov-Sinai entropy and the generalized Lyapunov exponent [5]. There, the value of the entropic index has been calculated analytically.

The explicit form of the *q*-exponential distribution is the following:

$$p_i = \frac{1}{\tilde{Z}_q(\lambda)} e_q(-\lambda Q_i) \qquad (i = 1, 2, \cdots, W), \qquad (1)$$

$$\tilde{Z}_q(\lambda) = \sum_{i=1}^{W} e_q(-\lambda Q_i), \qquad (2)$$

where $W$ is the number of accessible microscopic states of a system under consideration, $Q_i$ the *i*th value of a physical quantity $Q$, $\lambda$ a factor related to the

Lagrange multiplier, and $e_q(t)$ the $q$-exponential function defined by

$$e_q(t) = \begin{cases} [1+(1-q)t]^{1/(1-q)} & (1+(1-q)t > 0) \\ 0 & (1+(1-q)t \leq 0) \end{cases}. \qquad (3)$$

$q$ is a positive real number termed the *entropic index*. This distribution has the cut-off at $Q_{i,\max} = 1/[(1-q)\lambda]$ if $0 < q < 1$, whereas it is equivalent to the Zipf-Mandelbrot-type asymptotic power-law distribution with the exponent $1/(q-1)$ if $q > 1$. In the limit $q \to 1$, the $q$-exponential function converges to the ordinary exponential function and so does the $q$-exponential distribution to the Boltzmann-Gibbs-Jaynes exponential distribution.

Following Gibbs' procedure, one may also wish to derive the $q$-exponential distribution from the stationarity condition on a certain generalized entropy. Such an entropy is found to be not unique, however. There exist three known different entropies that are maximized by the $q$-exponential distribution under the constraint on the normalized $q$-expectation value of $Q$. This can be seen as follows.

Consider the functional

$$\Phi^{(J)}[p; \alpha, \beta] = S_q^{(J)}[p] - \alpha\left(\sum_{i=1}^{W} p_i - 1\right) - \beta\left(\sum_{i=1}^{W} P_i Q_i - Q_q\right)$$
$$(J = R, T, NT). \qquad (4)$$

$\alpha$ and $\beta$ are the Lagrange multipliers associated with the normalization condition on

the basic distribution, $\{p_i\}_{i=1, 2, \cdots, W}$, and the normalized $q$-expectation value of $Q$, $\sum_{i=1}^{W} P_i Q_i = Q_q$, where $P_i$ is the escort distribution [6] defined by

$$P_i = \frac{(p_i)^q}{\sum_{j=1}^{W}(p_j)^q}. \tag{5}$$

The three generalized entropies are listed as follows:

$$S_q^{(R)}[p] = \frac{1}{1-q} \ln \sum_{i=1}^{W} (p_i)^q, \tag{6}$$

$$S_q^{(T)}[p] = \frac{1}{1-q} \left[ \sum_{i=1}^{W} (p_i)^q - 1 \right], \tag{7}$$

$$S_q^{(NT)}[p] = \frac{1}{1-q} \left[ 1 - \frac{1}{\sum_{i=1}^{W} (p_i)^q} \right], \tag{8}$$

which are the Rényi entropy [7], the Tsallis entropy [8], and the normalized Tsallis entropy [9,10], respectively. These are connected to each other in the obvious ways, and all converges to the Boltzmann-Gibbs-Shannon entropy in the limit $q \to 1$:

$$\lim_{q \to 1} S_q^{(R)}[p] = \lim_{q \to 1} S_q^{(T)}[p] = \lim_{q \to 1} S_q^{(NT)}[p] = S[p] = -\sum_{i=1}^{W} p_i \ln p_i. \tag{9}$$

For a statistically independent bipartite system, $(A, B)$, these entropies satisfy

$$S_q^{(J)}(A, B) = S_q^{(J)}(A) + S_q^{(J)}(B) + \tau^{(J)}(q) S_q^{(J)}(A) S_q^{(J)}(B), \qquad (10)$$

where

$$\tau^{(R)}(q) = 0, \qquad (11)$$

$$\tau^{(T)}(q) = 1 - q, \qquad (12)$$

$$\tau^{(NT)}(q) = q - 1. \qquad (13)$$

Thus, the Rényi entropy is additive, whereas the Tsallis and normalized Tsallis entropies are nonadditive.

The Rényi entropy is conventionally used for the definition of the generalized dimension in multifractals [6], and the Tsallis entropy plays a central role in nonextensive statistical mechanics [2-4].

Variation of $\Phi^{(J)}$ with respect to $p_i$ gives rise to the following stationary distribution:

$$p_i^{(J)} = \frac{1}{\tilde{Z}_q^{(J)}(\lambda^{(J)})} e_q(-\lambda^{(J)} Q_i), \qquad (14)$$

$$\tilde{Z}_q^{(J)}(\lambda^{(J)}) = \sum_{i=1}^{W} e_q(-\lambda^{(J)} Q_i), \qquad (15)$$

where $\lambda$'s are given by

$$\lambda^{(R)} = \frac{\beta}{1 + (1-q)\beta Q_q^{(R)}}, \quad (16)$$

$$\lambda^{(T)} = \frac{\beta}{c_q^{(T)} + (1-q)\beta Q_q^{(T)}}, \quad (17)$$

$$\lambda^{(NT)} = \frac{\beta}{1/c_q^{(NT)} + (1-q)\beta Q_q^{(NT)}}, \quad (18)$$

respectively, provided that, in Eqs. (17) and (18), we have put

$$c_q^{(J)} = \sum_{i=1}^{W} (p_i^{(J)})^q. \quad (19)$$

Also, in the above expressions of $\lambda$'s, $Q_q^{(J)}$ stands for the normalized $q$-expectation value of $Q$ with respect to $p_i^{(J)}$ in Eq. (14). Here, it is worth mentioning that, as long as the $q$-exponential distribution is concerned, the expectation value has to be defined in terms of the escort distribution as in Eq. (4), since only in this case the principle of maximum generalized entropy can be consistent with the principle of equal *a priori* probability [11].

Thus, in fact, the Rényi, Tsallis, and normalized Tsallis entropies all lead to the $q$-exponential distributions of the same type. In other words, *mere fittings of observed data by the q-exponential distributions do not tell us anything about which the underlying physical entropy is*. In this respect, it should be noted that, in Ref. [12], Lesche has presented a counterexample showing instability of the Rényi entropy.

In this paper, we show that the Rényi and normalized Tsallis entropies are unstable under small deformation of a distribution and therefore cannot represent experimentally observable quantities, whereas the Tsallis entropy is stable and can provide an entropic basis for the $q$-exponential distributions. The discussion is general and is *independent of any stationary properties*.

The paper is organized as follows. In Sec. II, the rigorous definition of stability of a statistical quantity is given. In Sec. III, instability of the normalized Tsallis entropy as well as the Rényi entropy is shown. In Sec. IV, a general proof is established for stability of the Tsallis entropy. Sec. V is devoted to conclusion.

## II.    OBSERVABILITY AND STABILITY

Consider a statistical quantity $C = C[p]$, which has its maximum value, $C_{\max}$. $C[p]$ is said to be stable if the amount of its change under an arbitrary small deformation of the distribution remains small. Any observable quantities have to be stable, since otherwise their values cannot be experimentally reproducible. Let us measure the size of deformation from $\{p_i\}_{i=1,2,\cdots,W}$ to $\{p'_i\}_{i=1,2,\cdots,W}$ by the $l^1-$norm:

$$\| p - p' \|_1 = \sum_{i=1}^{W} | p_i - p'_i |. \tag{20}$$

Note that this quantity should be independent of *W*. Then, an observable quantity, $C[p]$, has to possess the following property [12]:

$$(\forall \varepsilon > 0) \ (\exists \delta > 0) \ \left( \| p - p' \|_1 \leq \delta \Rightarrow \left| \frac{C[p] - C[p']}{C_{\max}} \right| < \varepsilon \right) \tag{21}$$

for arbitrary values of *W*.

### III.    INSTABILITY OF RÉNYI AND NORMALIZED TSALLIS ENTROPIES IN THE THERMODYNAMIC LIMIT

In this section, we discuss instabilities of the Rényi and normalized Tsallis entropies by using a counterexample which violates the condition in Eq. (21).

First of all, we recall that Rényi, Tsallis, and normalized Tsallis entropies take their maximum values for the equiprobability $p_i = 1/W$ $(i = 1, 2, \cdots, W)$:

$$S_{q,\max}^{(R)} = \ln W, \tag{22}$$

$$S_{q,\max}^{(T)} = \ln_q W, \tag{23}$$

$$S_{q,\max}^{(NT)} = -\ln_q W^{-1}. \tag{24}$$

Here, $\ln_q x$ stands for the *q*-logarithmic function defined by

$$\ln_q x = \frac{1}{1-q}(x^{1-q} - 1) \qquad (x > 0) \tag{25}$$

which is the inverse function of the $q$-exponential function and converges to the ordinary logarithmic function in the limit $q \to 1$.

The deformation of a distribution to be examined is given as follows [12]:

- $0 < q < 1$;

$$p_i = \delta_{i1}, \quad p'_i = \left(1 - \frac{\delta}{2}\frac{W}{W-1}\right)p_i + \frac{\delta}{2}\frac{1}{W-1}. \tag{26}$$

- $q > 1$;

$$p_i = \frac{1}{W-1}(1 - \delta_{i1}), \quad p'_i = \left(1 - \frac{\delta}{2}\right)p_i + \frac{\delta}{2}\delta_{i1}. \tag{27}$$

Clearly this preserves the normalization condition. In both the cases of $0 < q < 1$ and $q > 1$, the $l^1$-norm is seen to be

$$\| p - p' \|_1 = \delta. \tag{28}$$

Also, from Eqs. (26) and (27), it is immediate to obtain

- $0 < q < 1$;

$$\sum_{i=1}^{W}(p_i)^q = 1, \quad \sum_{i=1}^{W}(p'_i)^q = \left(1 - \frac{\delta}{2}\right)^q + \left(\frac{\delta}{2}\right)^q (W-1)^{1-q}. \tag{29}$$

- $q > 1$;

$$\sum_{i=1}^{W} (p_i)^q = (W-1)^{1-q}, \qquad \sum_{i=1}^{W} (p'_i)^q = \left(\frac{\delta}{2}\right)^q + \left(1-\frac{\delta}{2}\right)^q (W-1)^{1-q}. \qquad (30)$$

### a.  Rényi entropy

The following discussion about instability of the Rényi entropy can be found in Ref. [12], but we present it here in order to make the discussion self-contained.

Using Eqs. (29) and (30) in Eq. (6), we find:

- $0 < q < 1$;

$$S_q^{(R)}[p] = 0, \quad S_q^{(R)}[p'] = \frac{1}{1-q} \ln\left[\left(1-\frac{\delta}{2}\right)^q + \left(\frac{\delta}{2}\right)^q (W-1)^{1-q}\right], \qquad (31)$$

$$\left|\frac{S_q^{(R)}[p] - S_q^{(R)}[p']}{S_{q,\max}^{(R)}}\right| = \left|\frac{\frac{1}{1-q}\ln\left[\left(1-\frac{\delta}{2}\right)^q + \left(\frac{\delta}{2}\right)^q (W-1)^{1-q}\right]}{\ln W}\right|$$

$$\to 1 \qquad (W \to \infty). \qquad (32)$$

- $q > 1$;

$$S_q^{(R)}[p] = \ln(W-1), \quad S_q^{(R)}[p'] = \frac{1}{1-q} \ln\left[\left(\frac{\delta}{2}\right)^q + \left(1-\frac{\delta}{2}\right)^q (W-1)^{1-q}\right], \qquad (33)$$

$$\left|\frac{S_q^{(R)}[p] - S_q^{(R)}[p']}{S_{q,\max}^{(R)}}\right| = \left|\frac{\ln(W-1) - \frac{1}{1-q}\ln\left[\left(\frac{\delta}{2}\right)^q + \left(1-\frac{\delta}{2}\right)^q (W-1)^{1-q}\right]}{\ln W}\right|$$

$$\to 1 \qquad (W \to \infty). \qquad (34)$$

Therefore, the condition in Eq. (21) is violated.

b. **Tsallis entropy**

Using Eqs. (29) and (30) in Eq. (7), we find:

- $0 < q < 1$;

$$S_q^{(T)}[p] = 0, \quad S_q^{(T)}[p'] = \frac{1}{1-q}\left[\left(1-\frac{\delta}{2}\right)^q + \left(\frac{\delta}{2}\right)^q (W-1)^{1-q} - 1\right], \quad (35)$$

$$\left|\frac{S_q^{(T)}[p] - S_q^{(T)}[p']}{S_{q,\max}^{(T)}}\right| = \left|\frac{\left(1-\frac{\delta}{2}\right)^q + \left(\frac{\delta}{2}\right)^q (W-1)^{1-q} - 1}{W^{1-q} - 1}\right|$$

$$\to \left(\frac{\delta}{2}\right)^q \quad (W \to \infty). \quad (36)$$

- $q > 1$;

$$S_q^{(T)}[p] = \ln_q(W-1), \quad S_q^{(T)}[p'] = \frac{1}{1-q}\left[\left(\frac{\delta}{2}\right)^q + \left(1-\frac{\delta}{2}\right)^q (W-1)^{1-q} - 1\right], \quad (37)$$

$$\left|\frac{S_q^{(T)}[p] - S_q^{(T)}[p']}{S_{q,\max}^{(T)}}\right| = \left|\frac{(W-1)^{1-q} - \left(\frac{\delta}{2}\right)^q - \left(1-\frac{\delta}{2}\right)^q (W-1)^{1-q}}{W^{1-q} - 1}\right|$$

$$\to \left(\frac{\delta}{2}\right)^q \quad (W \to \infty). \quad (38)$$

Therefore, if $\delta$ is taken to be $\delta < 2\varepsilon^{1/q}$, the condition in Eq. (21) is satisfied.

### c.  Normalized Tsallis entropy

Using Eqs. (29) and (30) in Eq. (8), we find:

- $0 < q < 1$;

$$S_q^{(NT)}[p] = 0, \quad S_q^{(NT)}[p'] = \frac{1}{1-q}\left[1 - \frac{1}{\left(1-\frac{\delta}{2}\right)^q + \left(\frac{\delta}{2}\right)^q (W-1)^{1-q}}\right], \tag{39}$$

$$\left|\frac{S_q^{(NT)}[p] - S_q^{(NT)}[p']}{S_{q,\max}^{(NT)}}\right| = \left|\frac{1 - \dfrac{1}{\left(1-\frac{\delta}{2}\right)^q + \left(\frac{\delta}{2}\right)^q (W-1)^{1-q}}}{1 - W^{q-1}}\right|$$

$$\to 1 \qquad (W \to \infty). \tag{40}$$

- $q > 1$;

$$S_q^{(NT)}[p] = -\ln_q (W-1)^{-1},$$

$$S_q^{(NT)}[p'] = \frac{1}{1-q}\left[1 - \frac{1}{\left(\frac{\delta}{2}\right)^q + \left(1-\frac{\delta}{2}\right)^q (W-1)^{1-q}}\right], \tag{41}$$

$$\left|\frac{S_q^{(NT)}[p] - S_q^{(NT)}[p']}{S_{q,\max}^{(NT)}}\right| = \left|\frac{-(W-1)^{q-1} + \dfrac{1}{\left(\frac{\delta}{2}\right)^q + \left(1-\frac{\delta}{2}\right)^q (W-1)^{1-q}}}{1 - W^{q-1}}\right|$$

$$\to 1 \qquad (W \to \infty). \tag{42}$$

Therefore, as in the case of the Rényi entropy, the condition in Eq. (21) is violated.

The results in Eqs. (32), (34), (40), and (42) mean that the Rényi and normalized Tsallis entropies with $0 < q < 1$ overestimate a large number of occupied states even if their overall probability is so small that they are irrelevant and those with $q > 1$ overestimate a high peak of probability.

Thus, among the three, there is a possibility only for the Tsallis entropy to be observable.

### IV. STABILITY OF TSALLIS ENTROPY

Stability of the Boltzmann-Gibbs-Shannon entropy has been shown in Ref. [12]. Here, we prove stability of the Tsallis entropy by generalizing the discussion in Ref. [12].

Let us define the following quantity:

$$A_q[p; t] = \sum_{i=1}^{W} \left( p_i - \frac{1}{e_q(t)} \right)_+, \quad (43)$$

where $t$ is a positive parameter and the symbol $(x)_+$ means

$$(x)_+ = \max\{x, 0\}. \quad (44)$$

The following will be useful later:

$$(x)_+ = x\theta(x), \tag{45}$$

$$\left|(x)_+ - (y)_+\right| \leq |x - y|, \tag{46}$$

where $\theta(x)$ is the Heaviside unit step function defined by $\theta(x) = 0$ for $x < 0$ and $\theta(x) = 1$ for $x > 0$.

The quantity in Eq. (43) has several interesting properties.

From Eq. (46), it immediately follows that

$$\left| A_q[p;t) - A_q[p';t) \right| \leq \|p - p'\|_1. \tag{47}$$

Using the relation

$$\left(\sum_{i=1}^{W}\left[p_i - \frac{1}{e_q(t)}\right]\right)_+ \leq \sum_{i=1}^{W}\left(p_i - \frac{1}{e_q(t)}\right)_+ < 1, \tag{48}$$

we have

$$\left(1 - \frac{W}{e_q(t)}\right)_+ \leq A_q[p;t) < 1. \tag{49}$$

In particular, if $t \geq \ln_q W$, then Eq. (49) becomes

$$1 - \frac{W}{e_q(t)} \leq A_q[p;t) < 1. \tag{50}$$

The same is true for another distribution $\{p'_i\}_{i=1,2,\cdots,W}$, that is,

$$-1 < -A_q[p';t] \leq -1 + \frac{W}{e_q(t)}. \tag{51}$$

Adding Eqs. (50) and (51), we find

$$\left| A_q[p;t] - A_q[p';t] \right| < \frac{W}{e_q(t)} \qquad (\forall t \geq \ln_q W). \tag{52}$$

In the limit $t \to t_{max}$ with $t_{max} = \infty$ $(0 < q < 1)$, $1/(q-1)$ $(q > 1)$ [see Eq. (3)], $e_q(t)$ diverges and therefore $A_q[p;t]$ tends to unity. So, the integral $\int_0^{t_{max}} dt \left(1 - A_q[p;t]\right)$ may converge. This is in fact the case. Using Eq. (45), this integral is written as

$$\int_0^{t_{max}} dt \left(1 - A_q[p;t]\right) = \sum_{i=1}^{W} \int_0^{t_{max}} dt \left( p_i - \frac{1}{e_q(t)} \right) \left[ 1 - \theta\left( p_i - \frac{1}{e_q(t)} \right) \right]$$
$$+ W \int_0^{t_{max}} dt \, \frac{1}{e_q(t)}. \tag{53}$$

The second term on the right-hand side gives

$$W \int_0^{t_{max}} dt \, \frac{1}{e_q(t)} = \frac{W}{q}. \tag{54}$$

On the other hand, noting $0 < \ln_q(1/p_i) < t_{max}$, the integral in the first term is

calculated as follows:

$$\sum_{i=1}^{W} \int_0^{t_{\max}} dt \left( p_i - \frac{1}{e_q(t)} \right) \left[ 1 - \theta\left( p_i - \frac{1}{e_q(t)} \right) \right]$$

$$= \sum_{i=1}^{W} \int_0^{\ln_q(1/p_i)} dt \left( p_i - \frac{1}{e_q(t)} \right)$$

$$= \frac{1}{1-q} \left[ \sum_{i=1}^{W} (p_i)^q - 1 \right] + \frac{1}{q} \sum_{i=1}^{W} (p_i)^q - \frac{W}{q}. \tag{55}$$

Therefore, we obtain

$$\int_0^{t_{\max}} dt \left( 1 - A_q[p;t] \right) = \frac{1}{q} S_q^{(T)}[p] + \frac{1}{q}. \tag{56}$$

or, conversely,

$$S_q^{(T)}[p] = -1 + q \int_0^{t_{\max}} dt \left( 1 - A_q[p;t] \right). \tag{57}$$

Now, using the representation in Eq. (57) and taking $a$ satisfying $-\ln_q W < 0 < a < t_{\max}$, we have

$$\left| S_q^{(T)}[p] - S_q^{(T)}[p'] \right| = q \left| \int_0^{t_{\max}} dt \left( A_q[p;t] - A_q[p';t] \right) \right|$$

$$\leq q \int_0^{t_{max}} dt \left| A_q[p;t] - A_q[p';t] \right|$$

$$= q \int_0^{a+\ln_q W} dt \left| A_q[p;t] - A_q[p';t] \right|$$

$$+ q \int_{a+\ln_q W}^{t_{max}} dt \left| A_q[p;t] - A_q[p';t] \right|. \tag{58}$$

From Eq. (47), the first integral is found to satisfy

$$\int_0^{a+\ln_q W} dt \left| A_q[p;t] - A_q[p';t] \right| \leq \| p - p' \|_1 (a + \ln_q W). \tag{59}$$

Likewise, from Eq. (52), the second integral is evaluated as

$$\int_{a+\ln_q W}^{t_{max}} dt \left| A_q[p;t] - A_q[p';t] \right| \leq \int_{a+\ln_q W}^{t_{max}} dt \frac{W}{e_q(t)} = \frac{W}{q} \left[ e_q(a + \ln_q W) \right]^{-q}. \tag{60}$$

Therefore, we have

$$\left| S_q^{(T)}[p] - S_q^{(T)}[p'] \right| \leq q \| p - p' \|_1 (a + \ln_q W) + \frac{W}{\left[ e_q(a + \ln_q W) \right]^q}. \tag{61}$$

This inequality holds for any values of $a$ satisfying $-\ln_q W < 0 < a < t_{max}$. Evaluating the minimum of the right-hand, which is realized when

$$a + \ln_q W = \ln_q \frac{W}{\| p - p' \|_1}, \tag{62}$$

Eq. (57) is reexpressed as follows:

$$\left| S_q^{(T)}[p] - S_q^{(T)}[p'] \right| \leq q \| p - p' \|_1 \ln_q \frac{W}{\| p - p' \|_1} + W^{1-q}(\| p - p' \|_1)^q. \tag{63}$$

Using the equality, $\ln_q(y/x) = x^{q-1}(\ln_q y - \ln_q x)$, we further obtain

$$\left| S_q^{(T)}[p] - S_q^{(T)}[p'] \right| \leq (\| p - p' \|_1)^q \ln_q W + (\| p - p' \|_1)^q (1 - q \ln_q \| p - p' \|_1), \tag{64}$$

from which we find

$$\left| \frac{S_q^{(T)}[p] - S_q^{(T)}[p']}{S_{q,\max}^{(T)}} \right| \leq \left| (\| p - p' \|_1)^q + \frac{(\| p - p' \|_1)^q (1 - q \ln_q \| p - p' \|_1)}{\ln_q W} \right|$$

$$\rightarrow \begin{cases} (\| p - p' \|_1)^q & (0 < q < 1) \\ q \| p - p' \|_1 & (q > 1) \end{cases} \quad (W \rightarrow \infty). \tag{65}$$

Therefore, taking $\| p - p' \|_1 \leq \delta < \varepsilon^{1/q}$ ($0 < q < 1$) or $\| p - p' \|_1 \leq \delta < \varepsilon / q$ ($q > 1$), we see that the condition in Eq. (21) is satisfied by the Tsallis entropy.

The above discussion holds for $\forall q > 0$, and so, as a simple byproduct, stability of the Boltzmann-Gibbs-Shannon entropy [12] corresponding to the limit $q \rightarrow 1$ (also of the Rényi and normalized Tsallis entropies) is reestablished.

## V. CONCLUSION

We have shown that among the Rényi, Tsallis, and normalized Tsallis entropies, only the Tsallis entropy is stable and can give rise to experimentally observable quantities. Therefore, it is the Tsallis entropy, on which the ubiquitous $q$-exponential distributions have their basis. A remaining important (and hard) question is if the Tsallis entropy is the unique generalized entropy. In this respect, we wish to mention that there are some affirmative points: there exist a set of axioms and the uniqueness theorem for the Tsallis entropy [13], and the structure of nonadditivity in Eq. (10) is essential from the viewpoint of the zeroth law of thermodynamics [14,15].